\documentclass[twocolumn,aps,superscriptaddress,prb]{revtex4-1}
\usepackage{graphicx}
\usepackage{amssymb}
\usepackage{amsmath}
\usepackage{bm}
\usepackage{color}

\def\Journal #1,#2,#3,#4#5#6#7{#1 {\bf #2}, #3 (#4#5#6#7)}

\def\Vec{\mathbf}
\def\GVec#1{\mbox{\boldmath $#1$}}

\def\bc{{\bf c}}

\begin{document}

\title{Maximally-localized Wannier orbitals and the extended Hubbard model for the twisted bilayer graphene\\
}
\author{Mikito Koshino}
\thanks{koshino@phys.sci.osaka-u.ac.jp}
\affiliation{Department of Physics, Osaka University,  Toyonaka 560-0043, Japan}

\author{Noah F. Q. Yuan}
\affiliation{Department of Physics, Massachusetts Institute of Technology, Cambridge, Massachusetts 02139, USA}

\author{Takashi Koretsune}
\affiliation{Department of Physics, Tohoku University,  Sendai 980-8578, Japan}

\author{Masayuki Ochi}
\affiliation{Department of Physics, Osaka University,  Toyonaka 560-0043, Japan}

\author{Kazuhiko Kuroki}
\affiliation{Department of Physics, Osaka University,  Toyonaka 560-0043, Japan}

\author{Liang Fu}
\affiliation{Department of Physics, Massachusetts Institute of Technology, Cambridge, Massachusetts 02139, USA}

\begin{abstract}
We develop an effective extended Hubbard model to describe the low-energy electronic properties 
of the twisted bilayer graphene.
By using the Bloch states in the effective continuum model
and with the aid of the maximally localized algorithm,
we construct the Wannier orbitals and obtain an effective tight-binding model on the emergent honeycomb lattice.
We found the Wannier state takes a peculiar three-peak form 
in which the amplitude maxima are located at the triangle corners surrounding the center.
We estimate the direct Coulomb interaction and the exchange interaction between the Wannier states.
At the filling of two electrons per super cell, in particular, we find 
an unexpected coincidence in the direct Coulomb energy between
a charge-ordered state and a homogeneous state, which would possibly lead to an unconventional many-body state.
\end{abstract}

\maketitle

\section{Introduction}

The recent discovery of the superconductivity and strongly correlated insulating state 
in the twisted bilayer graphene (TBG) \cite{cao2018unconventional,cao2018mott} attracts enormous attention and 
triggered a surge of theoretical works on this subject.\cite{xu2018topological,yuan2018model,po2018origin,roy2018unconventional,guo2018pairing,padhi2018wigner,dodaro2018phases,huang2018antiferromagnetically,zhang2018low,ray2018wannier,liu2018chiral,xu2018kekul,peltonen2018mean,fidrysiak2018unconventional}
TBG is a bilayer system in which two graphene layers are rotationally stacked to each other,\cite{berger2006electronic,hass2007structural,hass2008multilayer,li2009observation,miller2010structural,luican2011single}where the electronic band structure sensitively depends on the twist angle $\theta$. 
In a small $\theta$, in particular, a slight difference in the lattice orientation 
gives rise to a long-period moir\'{e} interference pattern,
causing a substantial modification of the Dirac dispersion.\cite{lopes2007graphene,mele2010commensuration,trambly2010localization,shallcross2010electronic,morell2010flat,bistritzer2011moirepnas,kindermann2011local,xian2011effects,PhysRevB.86.155449,moon2012energy,de2012numerical,moon2013opticalabsorption}. 
Theoretically it was predicted that
the Fermi velocity vanishes at certain $\theta$'s called the magic angles,
around which nearly-flat bands with extremely narrow band width emerge at low energy. \cite{bistritzer2011moirepnas,de2012numerical}
The superconductivity was actually observed around a magic angle of $1.05^\circ$,
where the insulating phase and the superconducting phase occur around the filling of two electrons per super cell.

It is a challenging problem to theoretically describe the many-body physics in TBG.
At $\theta =1.05^\circ$, the spatial period of the moir\'{e} pattern is more than 10 nm 
and the number of carbon atoms in a unit cell exceeds 10,000.
The electronic property of such a huge and complex system can be calculated efficiently
by the effective continuum model which captures the long-wavelength physics associated with the moir\'{e} period.
\cite{lopes2007graphene,bistritzer2011moirepnas,kindermann2011local,PhysRevB.86.155449,moon2013opticalabsorption,koshino2015interlayer,koshino2015electronic}
However, the effective continuum energy spectrum still contains a number of energy bands in the low-energy region,
and we need one more step to simplify the model so as to exclusively describe the nearly-flat bands at lowest energy.

Actually the nearly-flat bands are separated by the energy gaps from other bands, \cite{cao2016superlattice,cao2018unconventional,nam2017lattice}
making it possible to construct an effective lattice model
with well-localized Wannier orbitals purely consisting of the flat band states.
Such an effective model was actually predicted by the symmetry analysis,\cite{yuan2018model}
which concludes that the Wannier orbitals should be centered at non-equivalent AB spot and BA spot in 
the moir\'{e} pattern, to form an emergent honeycomb lattice. Arguments and calculations suggesting a honeycomb lattice description have also been put forward in an independent work.\cite{po2018origin} 
To obtain a concrete model with specific parameters, 
we need to construct the actual Wannier orbitals from the realistic model of TBG.

In this paper, we develop an extended Hubbard model of TBG at the magic-angle ($\theta =1.05^\circ$),
based on the effective continuum model.
By taking an appropriate linear combination of the Bloch states in the nearly flat bands,
we build the Wannier orbitals centered at AB and BA spots,
and obtain the effective tight-binding model on the emergent honeycomb lattice.
Here we adopted the maximally localized algorithm \cite{marzari1997maximally}
to minimize the spread of the wave functions.
From the obtained Wannier orbitals,  we estimate the direct 
Coulomb energy and the exchange energy between electrons residing at different orbitals.
The obtained Wannier state is centered at AB or BA spot,
while its maximum amplitude is found to be not at the center, 
but at three AA spots surrounding the center, as also noticed in Ref.\ \onlinecite{po2018origin}. 
Importantly, the pair of Wannier orbitals that we constructed explicitly has 
$(p_x, p_y)$ on-site symmetry, 
hence forms a doublet under three-fold rotation around their centers, consistent 
with the symmetry analysis.\cite{yuan2018model}

Due to this peculiar three-peak form,
the electron-electron interaction between the neighboring sites is as important as the on-site interaction term.
At the filling of two electrons per super cell, 
in particular, we find 
an unexpected coincidence in the direct Coulomb energy between
two different many-body states:
a homogeneous state where an electron enters every sublattice of the effective honeycomb lattice,
and a charge-ordered state where two electrons reside at every two sublattices [Fig.\ \ref{fig_manybody}(a) and (b)].
We expect that such competing nature would possibly give rise to a nontrivial many-body ground state.

This paper is organized as follows: In Sec.\ \ref{sec_atomic}, we explain the atomic structure of TBG,
and in Sec.\ \ref{sec_effective}, we introduce the effective continuum model and argue the
structure of the nearly-flat bands at the magic angle $\theta =1.05^\circ$.
In Sec.\ \ref{sec_wannier}, we construct the Wannier orbitals using the maximally localizing method,
and obtain the tight-binding model in Sec.\ \ref{sec_tight-binding}.
We consider the electron-electron interaction between the Wannier states in Sec.\ \ref{sec_interaction}.
A brief conclusion is presented in Sec.\ \ref{sec_conclusion}.


\section{Atomic structure}
\label{sec_atomic}

We define the atomic structure of TBG 
by starting from AA-stacked bilayer graphene (i.e. perfectly overlapping honeycomb lattices)
and rotating the layer 1 and 2 around a pair of registered $B$-sites by $-\theta/2$ and $+\theta/2$, respectively.
We define  $\Vec{a}_1 = a(1,0)$ and $\Vec{a}_2 = a(1/2,\sqrt{3}/2)$ 
as the lattice vectors of the initial AA-stacked bilayer
before the rotation, where $a \approx 0.246\,\mathrm{nm}$ is the lattice constant of graphene.
The corresponding reciprocal lattice vectors 
are $\Vec{a}^*_1 = (2\pi/a)(1,-1/\sqrt{3})$ and $\Vec{a}^*_2=(2\pi/a)(0,2/\sqrt{3})$.
After the rotation, the lattice vectors of layer $l$ are given by $\Vec{a}_i^{(l)} =R(\mp \theta/2)\Vec{a}_i$ 
with $\mp$ for $l=1,2$, 
respectively, where $R(\theta)$ represents the rotation by $\theta$.
Likewise, the reciprocal lattice vectors become $\Vec{a}_i^{*(l)} =R(\mp \theta/2)\Vec{a}^*_i$. With respect to the registered $B$-sites, TBG has point group $D_3$ generated by a three-fold in-plane rotation $C_{3z}$ along $z$-axis and a two-fold rotation $C_{2y}$ along $y$-axis.

\begin{figure}
\begin{center}
\leavevmode\includegraphics[width=0.9\hsize]{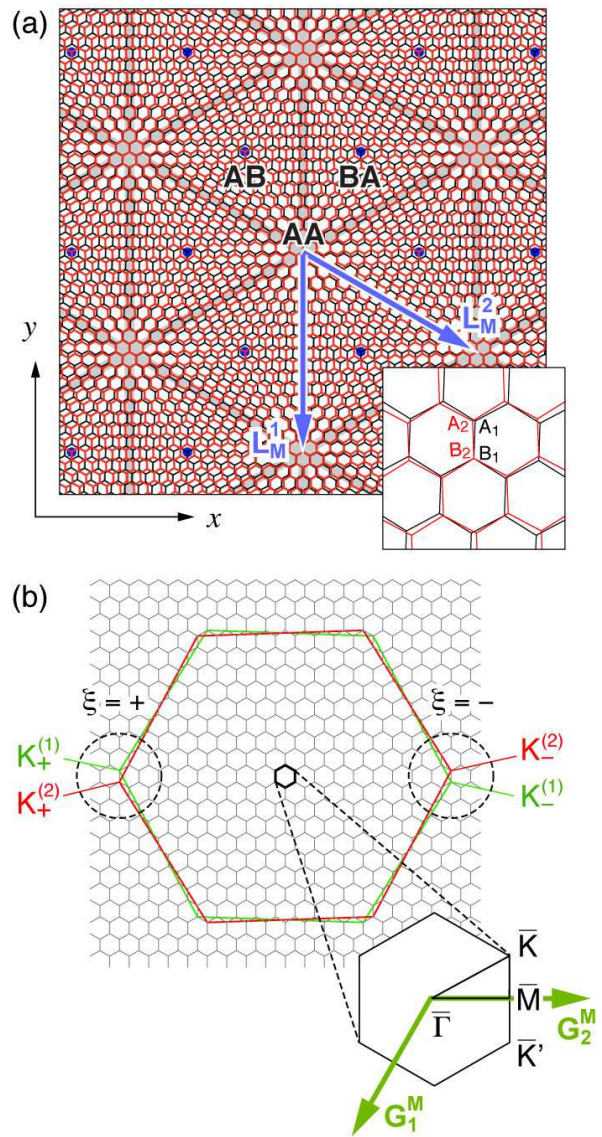}
\end{center}
\caption{
(a) Atomic structure of TBG with $\theta = 3.89^\circ$ and $D_3$ symmetry.
AA spots are located at the crossing points of the grid lines,
and AB and BA spots are at the centers of triangles indicated by dots.
(b) Brillouin zone folding in TBG with $\theta = 3.89^\circ$.
Two large hexagons represent the first Brillouin zones of graphene layer 1 and 2, 
and the small hexagon is the moir\'{e} Brillouin zone of TBG.
}
\label{fig_lattice_BZ}
\end{figure}

In a small angle TBG, the slight mismatch of the lattice periods of two layers gives rise to a long- period moir\'{e} 
interference pattern. The reciprocal lattice vectors for the moir\'{e} pattern is given by
$ \Vec{G}^{\rm M}_{i}  =  \textbf{a}^{*(1)}_i - \textbf{a}^{*(2)}_i \, (i=1,2).$
The real-space lattice vectors $\Vec{L}^{\rm M}_{j}$ can then be obtained from
$\Vec{G}^{\rm M}_i\cdot\Vec{L}^{\rm M}_{j} = 2\pi\delta_{ij}$.
A moir\'{e} unit cell is spanned by $\Vec{L}^{\rm M}_{1}$ and $\Vec{L}^{\rm M}_2$.
The lattice constant $L_{\rm M} = | \Vec{L}^{\rm M}_{1}|=| \Vec{L}^{\rm M}_2|$
is $L_{\rm M} =  a/[2\sin (\theta/2)]$.
Figure \ref{fig_lattice_BZ}(a) illustrates the atomic structure
of TBG with $\theta = 3.89^\circ$.
The lattice structure locally resembles the regular stacking such as AA, AB or BA depending on the position,
where AA represents the perfect overlapping of hexagons,
and AB (BA) is the shifted configuration in which A$_1$(B$_1$) sublattice is right above B$_2$(A$_2$).
In Fig.\ \ref{fig_lattice_BZ}(a), AA spots are located at the crossing points of the grid lines,
and AB and BA spots are at the centers of triangles indicated by dots.
Figure \ref{fig_lattice_BZ}(b) shows the corresponding folding of the Brillouin zone, 
where two large hexagons represent the first Brillouin zones of layer 1 and 2, 
and the small hexagon is the moir\'{e} Brillouin zone of TBG.
The graphene's Dirac points (the band touching points) 
are located  at $\Vec{K}^{(l)}_\xi = -\xi [2\Vec{a}^{(l)*}_1+\Vec{a}^{(l)*}_2]/3$ 
for layer $l$, where $\xi=\pm 1$ is the valley index.
We label the symmetric points of the reduced Brillouin zone as
$\bar{\Gamma}$, $\bar{M}$, $\bar{K}$ and $\bar{K'}$ as in Fig.\ \ref{fig_lattice_BZ}(b).

We can construct the TBG in alternative manners, for example, by rotating around the hexagon centers instead of the B site.
In that case, we have the different superlattice structure with point group $D_6$.
For completeness, we leave the discussion of $D_6$ structure and other superlattice structures in the Supplementary Material.\cite{suppl1}

\begin{figure}
\begin{center}
\leavevmode\includegraphics[width=1.\hsize]{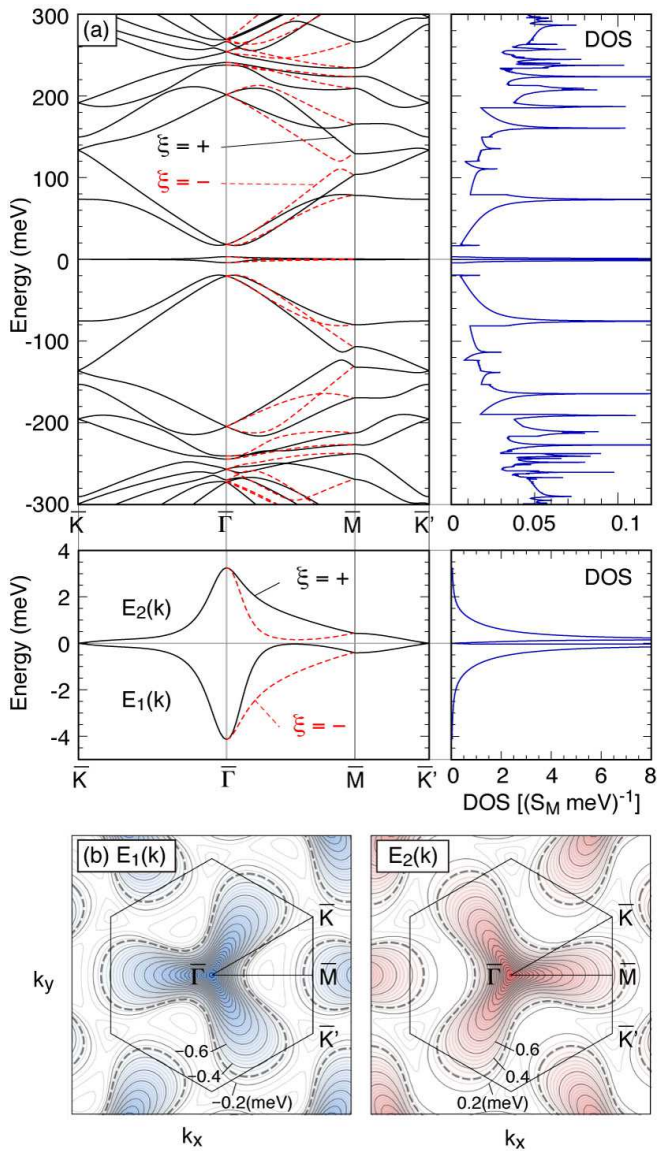}
\end{center}
\caption{
(a) Energy band and the density of states of TBG at $\theta=1.05^\circ$,
where the lower panel is the enlarged plot of the zero-energy region.
The black solid line and red dashed line represent the energy bands of $\xi =\pm$ valleys, respectively.
(b) Contour plots of  $E_1(\Vec{k})$ and $E_2(\Vec{k})$ for the valley $\xi = +$. 
The dashed contour corresponds to the filling of two electrons / holes per super cell ($n/n_0=\pm 2$).
}
\label{fig_band}
\end{figure}

\section{Effective continuum model}
\label{sec_effective}

When the moir\'{e}  period is much longer than the atomic scale, 
the electronic structure can be described by an effective continuum model.
\cite{lopes2007graphene,bistritzer2011moirepnas,kindermann2011local,
PhysRevB.86.155449,moon2013opticalabsorption,koshino2015interlayer,koshino2015electronic}
There the intervalley mixing between $\xi = \pm$ can be safely neglected
and the total Hamiltonian is block-diagonalized into the two independent valleys.
 The effective Hamiltonian of continuum model for the valley $\xi$ is written in a 4 $\times $ 4 matrix for the basis of $(A_1,B_1,A_2,B_2)$ as
 \begin{eqnarray}
	{H}^{(\xi)} = 
	\begin{pmatrix}
		H_1 & U^\dagger \\
		U & H_2
	\end{pmatrix}.
	\label{eq_H_eff}
\end{eqnarray}
Here $H_l (l=1,2)$ is the intralayer Hamiltonian of layer $l$,
which is given by the two-dimensional Weyl equation centered at $\Vec{K}^{(l)}_\xi$ point, 
\begin{align}
& H_l = - \hbar v [R(\pm \theta/2)({\Vec{k}}-\Vec{K}^{(l)}_\xi)] \cdot (\xi \sigma_x, \sigma_y),
	\label{eq_intra}
\end{align}
where $\pm$ is for $l=1$ and 2, respectively. We take $\hbar v /a = 2.1354$ eV.\cite{moon2013opticalabsorption}
$U$ is the effective interlayer coupling given by \cite{moon2013opticalabsorption,koshino2015interlayer,koshino2015electronic}
\begin{align}
 U &= 
\begin{pmatrix}
U_{A_2 A_1} & U_{A_2 B_1}
\\
U_{B_2 A_1} & U_{B_2 B_1}
\end{pmatrix}
\nonumber\\
&
=
\begin{pmatrix}
u & u'
\\
u' & u
\end{pmatrix}
+
\begin{pmatrix}
u & u'\omega^{-\xi}
\\
u'\omega^\xi & u
\end{pmatrix}
e^{i\xi \Vec{G}^{\rm M}_1\cdot\Vec{r}}
\nonumber\\
& 
\qquad \qquad +
\begin{pmatrix}
u & u'\omega^\xi
\\
u'\omega^{-\xi} & u
\end{pmatrix}
e^{i\xi(\Vec{G}^{\rm M}_1+\Vec{G}^{\rm M}_2)\cdot\Vec{r}},
\label{eq_interlayer_matrix}
\end{align}
where $\omega =e^{2\pi i/3}$. 
Here $u$ and $u'$ describe the amplitudes of diagonal and off-diagonal terms, respectively, in the sublattice space.
The effective models in the previous studies \cite{moon2013opticalabsorption,koshino2015interlayer,koshino2015electronic} assume $u=u'$,
which corresponds to a flat TBG in which the interlayer spacing $d$ is constant everywhere.
On the other hand, several theoretical studies predicted that the optimized lattice structure of TBG is actually corrugated
in the out-of-plane direction, in such a way that
$d$ is the widest in AA stacking region and the narrowest AB / BA stacking region. \cite{uchida2014atomic,van2015relaxation,dai2016twisted, jain2016structure}
Here we incorporate the corrugation effect as a  difference between $u = 0.0797$eV and $u' = 0.0975$eV in the effective model,
of which detailed derivation is presented in the Appendix \ref{sec_app}.
As we show in the following, the difference between $u$ and $u'$ introduces energy gaps between
the lowest bands and the excited bands, 
in a qualitative agreement with the experimental observation.\cite{cao2016superlattice,cao2018unconventional,cao2018mott}
It was found that the energy gaps isolating the lowest nearly-flat bands are also caused by the in-plane distortion.\cite{nam2017lattice}

 The calculation of the energy bands  and the eigenstates 
 is done in the $k$-space picture.
For a single Bloch vector $\Vec{k}$ in the moir\'{e} Brillouin zone,   
the moir\'{e} interlayer coupling hybridizes the graphene's eigenstates at
$\Vec{q} = \Vec{k} + \Vec{G}$, where
$\Vec{G} = m_1  \Vec{G}^{\rm M}_1 + m_2  \Vec{G}^{\rm M}_2$
and $m_1$ and $m_2$ are integers.
Therefore the eigenstate is written as
\begin{equation}
\psi^{X}_{n\Vec{k}}(\Vec{r})
 = \sum_{\Vec{G}}
C^{X}_{n\Vec{k}}(\Vec{G})\\[0.2em]
e^{i (\Vec{k}+\Vec{G})\cdot\Vec{r}},
\label{eq_bloch_wave}
\end{equation}
where $X=A_1,B_1,A_2,B_2$ is the sublattice index, 
$n$ is the band index and $\Vec{k}$ is the Bloch wave vector in the moir\'{e} Brillouin zone.
As the low-energy states are expected to be dominated by the individual graphenes' eigenstates near the original Dirac points,
we pick up $\Vec{q}$'s inside the cut-off circle $|\Vec{q}-\Vec{q}_0| < q_c$,
where $\Vec{q}_0$ is taken as the midpoint between $\Vec{K}^{(1)}_\xi$ and $\Vec{K}^{(2)}_\xi$,
and $q_c$ is set to $4 G_{\rm M} \, (G_{\rm M} = |\Vec{G}^{\rm M}_1|= |\Vec{G}^{\rm M}_2|)$.
Since the intervalley coupling can be neglected, the calculation is done independently for each of $\xi=\pm$ as we discussed previously.
We then numerically diagonalize the Hamiltonian within the limited wave space inside the cut-off circle
and obtain the eigenenergies and eigenstates.


Figure \ref{fig_band}(a) shows the energy band and the density of states
of TBG at the magic angle $\theta=1.05^\circ$, calculated by this approach.
Here in the following, the origin of band energy axis is set to the charge neutral point.
The lower panel is the enlarged plot of the zero-energy region where the near-flat bands are located.
The black solid line and red dashed line represent the energy bands of $\xi =\pm$ valleys, respectively.
They are the time-reversal partners to each other, and 
the energy bands of $\xi =-$ are obtained just by inverting $\Vec{k}$ to
$-\Vec{k}$. The flat band cluster consists of two bands 
per spin and valley, which are denoted as $E_1(\Vec{k})$ and $E_2(\Vec{k})$ for 
the hole side and the electron side, respectively.
The overall structure is about 7.5 meV wide in energy axis,
and separated from the excited bands by the energy gap of about 14 meV
in each of the electron side and the hole side.
Figure \ref{fig_band}(b) shows the contour plots of  $E_1(\Vec{k})$ and $E_2(\Vec{k})$
for the valley $\xi = +$. 
$E_1(\Vec{k})$ and $E_2(\Vec{k})$ are trigonally warped in the opposite directions,
so that $E_1(\Vec{k})\neq E_1(-\Vec{k})$ and $E_2(\Vec{k})\neq E_2(-\Vec{k})$.
The particle-hole symmetry is absent and the $E_1$ band is wider than the $E_2$ band.
The van-Hove singularity is located at $E\approx  -0.11$ meV and 0.16 meV,
which correspond to the carrier density $n/n_0\approx - 0.78$ and 0.63, respectively, with spin and valley included.
Here $n_0 = 1/S_{\rm M}$, $S_{\rm M} = (\sqrt{3}/2)L_{\rm M}^2$ is the moir\'{e} unit area
(the band gap is $n/n_0=\pm 4$) and $L_{\rm M}$ is 13.4nm at $\theta=1.05^\circ$.
The filling of two electrons / holes per super cell ($n/n_0=\pm 2$)
corresponds to $E\approx  0.289$ meV and $-0.286$ meV, respectively, which are indicated by dashed contours in
Fig. \ref{fig_band}(b).

\begin{figure*}
\begin{center}
\leavevmode\includegraphics[width=1.\hsize]{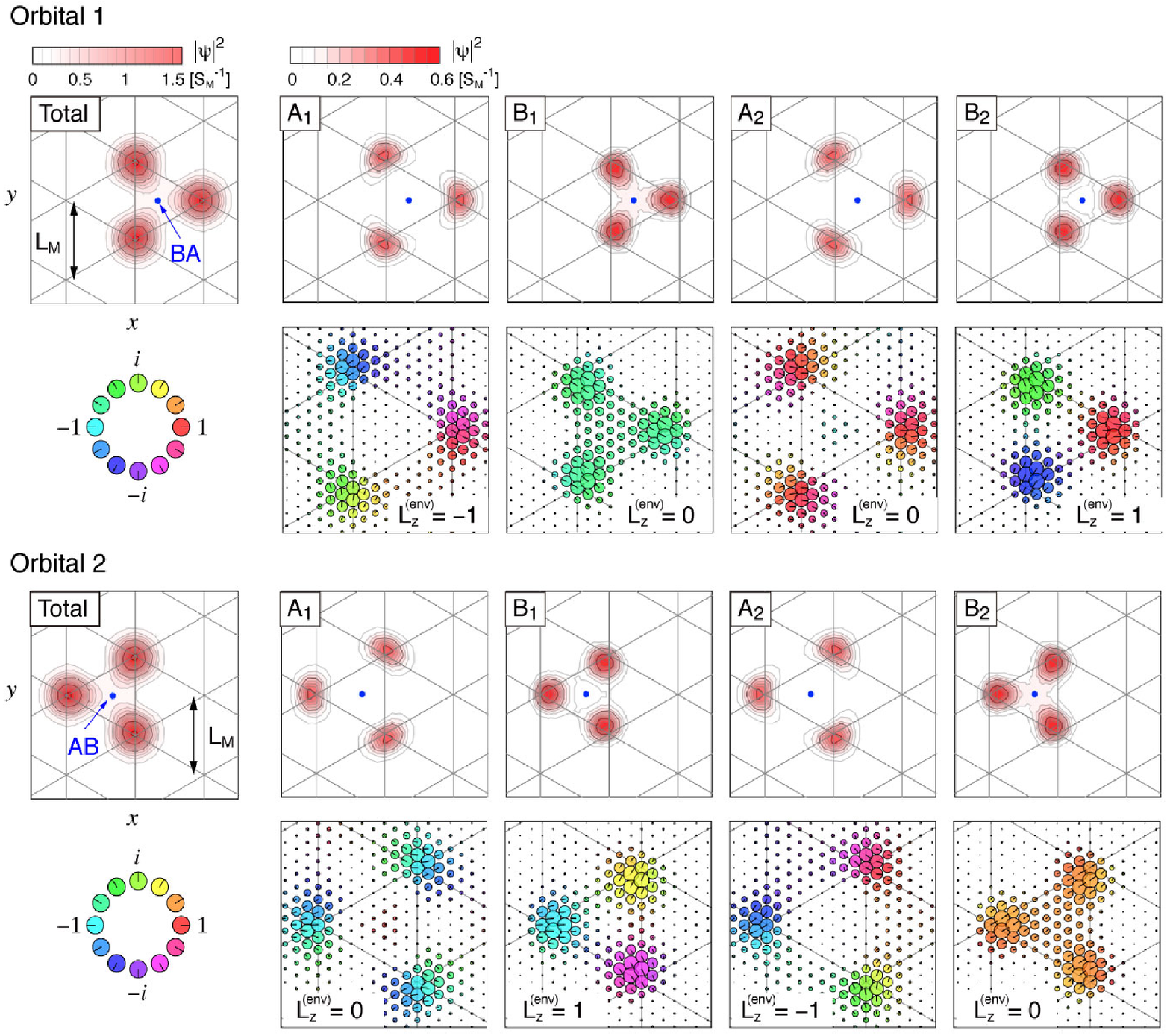}
\end{center}
\caption{
Maximally localized Wannier orbitals of the valley $\xi=+$,
in the low-energy flat band of TBG with $\theta=1.05^\circ$.
For each of orbital 1 and 2, the top five panels show the contour maps for 
the squared amplitudes of the total wave function and of the four sublattice components.
The lower panels illustrate the phase of the envelope function on some sample points, where
the amplitude is indicated by the radius of a circle, and its phase factor is by the direction of a bar and also by color.
}
\label{fig_wannier}
\end{figure*}

\section{Wannier orbitals}
\label{sec_wannier}

We construct the localized Wannier orbitals from the Bloch wave functions of the effective model.
Since the nearly flat bands are energetically isolated from other bands,
we expect that well-localized orbits can be made 
purely from the flat band states with all other bands neglected.
The number of the independent Wannier orbitals in a unit cell
coincides with the number of the energy bands taken into account, 
so we have two Wannier orbitals per spin and valley.
According to the symmetry analysis \cite{yuan2018model},
the two orbitals should be centered at AB and BA spots to form a honeycomb lattice.
Our strategy is to first prepare certain initial orbitals centered at AB and BA,
and then apply the maximally localized algorithm. \cite{marzari1997maximally}
The following process is applied to $\xi=\pm$ valleys separately,
and we omit the valley index $\xi$ hereafter.

The initial wave functions can be prepared as follows.
First we fix the global phase factor of the Bloch states in two different ways:
In gauge 1, we fix the phase so that $\psi^{B_1}_{n\Vec{k}}(\Vec{r}_{\rm BA})$ is real,
and in gauge 2, we fix the phase so that $\psi^{A_1}_{n\Vec{k}}(\Vec{r}_{\rm AB})$ is real.
Here $\Vec{r}_{\rm BA} = (1/2, \sqrt{3}/2)(L_{\rm M}/\sqrt{3})$ and $\Vec{r}_{\rm AB}= (-1/2, \sqrt{3}/2)(L_{\rm M}/\sqrt{3})$ 
are the positions of BA and AB spots, respectively, 
measured from the AA spot $(0,0)$ [Fig.\ \ref{fig_lattice_BZ}(a)].
We write the Bloch function in the gauge 1 as $\psi_{n\Vec{k}}$,
and that in the gauge 2 as  $e^{i\phi_{n\Vec{k}}}\psi_{n\Vec{k}}$, 
where $e^{i\phi_{n\Vec{k}}}$ is the relative phase factor between gauge 1 and 2.
We construct the initial Wannier orbitals 1 and 2 by summing the Bloch states 
of the bands $\psi_{1\Vec{k}}$ and $\psi_{2\Vec{k}}$ 
(corresponding to $E_1(\Vec{k})$ and $E_2(\Vec{k})$, respectively) as,
\begin{align}
& | \Vec{R}, 1\rangle_0 = \frac{1}{\sqrt{N}}\sum_{\Vec{k}} e^{-i\Vec{k} \cdot \Vec{R}}
\frac{1}{\sqrt{2}} (|\psi_{1\Vec{k}}\rangle + |\psi_{2\Vec{k}}\rangle) \nonumber\\
 & | \Vec{R}, 2\rangle_0  = \frac{1}{\sqrt{N}}\sum_{\Vec{k}} e^{-i\Vec{k} \cdot \Vec{R}} 
\frac{1}{\sqrt{2}} e^{i\phi_{1\Vec{k}}} (|\psi_{1\Vec{k}}\rangle - |\psi_{2\Vec{k}}\rangle),
 \label{eq_wannier_ini}
\end{align}
Here $\Vec{R}=n_1\Vec{L}_1^{\rm M}+n_2\Vec{L}_2^{\rm M}$ is the moir\'{e} lattice vector,
and the summation in $\Vec{k}$ is taken over $N$ discrete points in the moir\'{e} Brillouin zone.
We take $N = 18\times 18$ in this study.
It is straightforward to check the orthonormality,
$_0\langle  \Vec{R}', n' | \Vec{R}, n\rangle_0 = \delta_{\Vec{R},\Vec{R}'}\delta_{n,n'}$.

While $|\Vec{R}, 1\rangle_0$ and $| \Vec{R}, 2\rangle_0$
are already well localized around the center positions 
$\Vec{R}+\Vec{r}_{\rm BA}$ and $\Vec{R}+\Vec{r}_{\rm AB}$, respectively,
we can even reduce the spread of the wave function by maximally localizing method \cite{marzari1997maximally}.
The final expression for the orbital $n(=1,2)$ is given by
\begin{align}
& | \Vec{R}, n\rangle= \frac{1}{\sqrt{N}}\sum_{\Vec{k}} e^{-i\Vec{k} \cdot \Vec{R}} 
\sum_{m=1,2} U^{(\Vec{k})}_{mn} |\psi_{m\Vec{k}}\rangle,
 \end{align}
where $U^{(\Vec{k})}_{mn}$ is a $2\times 2$ unitary matrix. 
The algorithm optimizes $U^{(\Vec{k})}_{mn}$ to minimize the spread functional.
We put Eq.\ (\ref{eq_wannier_ini}) as the initial value of $U^{(\Vec{k})}_{mn}$, 
and iterate the minimization process until the convergence.
In each step, we impose the symmetry constraint to $U^{(\Vec{k})}_{mn}$.
The optimized Wannier orbitals for the valley $\xi=+$ are illustrated in Fig.\ \ref{fig_wannier}.
Those for the opposite valley $\xi=-$ are given by the complex conjugate.
For each of orbital 1 and 2, the top five panels show the contour maps for 
the squared amplitudes of the total wave function and of the four sublattice components.
We actually see that the orbital 1 and 2 are centered at BA and AB positions, respectively, while
the maximum of the wave amplitudes are located not at the center, but near three AA spots surrounding the center.
This reflects the fact that the Bloch wave functions of the nearly-flat bands are mostly localized AA spot of the moir\'{e}
pattern.\cite{trambly2010localization,uchida2014atomic}

The lower panels illustrate the phase of the envelope function $F^{X_l}(\Vec{r})$ ($X=A,B$ and $l=1,2$) 
on some sample points,
where the total wave function is $\psi^{X_l}(\Vec{r}) = e^{i \Vec{K}^{(l)}_\xi \cdot \Vec{r}}F^{X_l}(\Vec{r})$.
Here the absolute value of $F^{X_l}(\Vec{r})$ is indicated by
the radius of a circle, and its phase factor is by the direction of a bar and also by color.
Now we see that the envelope functions on different sublattices have different eigenvalues of $C'_{3z}$,
in-plane rotation with respect to its own center.
However, noting that the Bloch factor $e^{i \Vec{K}_\xi \cdot \Vec{r}}$ also carries a non-zero eigenvalue of $C'_{3z}$,
the total wave function $\psi=(\psi^{A_1},\psi^{B_1},\psi^{A_2},\psi^{B_2})$ is found to be
an eigenstate of $C'_{3z}$ with a single eigenvalue.
In orbital 1, for example, the $C'_{3z}$ eigenvalue of $F^{X_l}$ is $(\omega, 1, 1, \omega^*)$ 
for $(A_1, B_1, A_2, B_2)$, so that the angular momentum of the envelope function
is written as $L_z^{\rm (env)} = (-1, 0, 0, 1)$.
On the other hand, the $C'_{3z}$ eigenvalue for the Bloch factor $e^{i \Vec{K}_\xi \cdot \Vec{r}}$ can be found 
by noting that BA spot (the orbital center) coincides with $A_1$ site and the center of hexagon of layer 2 
[Fig.\ \ref{fig_lattice_BZ}(a)],
and then we obtain $L_z^{\rm (Bloch)} = (0, -1, -1, 1)$.
Therefore, the total angular momentum  $L_z = L_z^{\rm (env)} + L_z^{\rm (Bloch)}$ is $-1$ for all the sublattices.
Similarly, we can show $L_z=-1$ also for orbital 2.
Since the Wannier functions at the opposite valleys are related by the complex conjugate,
we finally conclude that the eigenvalue of $C'_{3z}$ is $\omega^{\xi} = e^{\xi 2\pi i/3}$ for both orbital 1 and 2.
Namely orbital 1 and 2 from the same valley $ \xi $ have the same nonzero angular momentum $ L_{z}=-\xi$,
in accordance with the symmetry analysis.\cite{yuan2018model}

The initial guess of the Wannier orbital in Eq.\ (\ref{eq_wannier_ini})
is closely related to the angular momentum of the envelope function. 
For the orbital 1, the envelope function of $B_1$ has zero angular momentum,
so that it has a finite amplitude at the orbital center $\Vec{r}_{\rm BA}$ as seen in Fig.\ \ref{fig_wannier}.
It does not contradict with the nonzero total angular momentum $L_z=-1$, because
BA spot coincides with the hexagon center of layer 1, but not $B_1$ site.
The finite amplitude at $\Vec{r}_{\rm BA}$ is actually linked to the gauge choice for $| \Vec{R}, 1\rangle_0$,
which requires that $\psi^{B_1}_{n\Vec{k}}(\Vec{r}_{\rm BA})$ is real.
There all the wave functions add up in the same phase at $\Vec{r}_{\rm BA}$, 
so that we have an orbital localized at $\Vec{r}_{\rm BA}$ with finite amplitude.
The same is true for the orbital 2, of which envelope angular momentum vanishes at $A_1$.
The wrong gauge choices (e.g., $\psi^{A_1}_{n\Vec{k}}(\Vec{r}_{\rm BA})$ is real)
do not make a well localized orbital, because the angular momentum of the Wannier function is forced by the symmetry. 
Also, the hybridized form of $|\psi_{1\Vec{k}}\rangle \pm |\psi_{2\Vec{k}}\rangle$  in Eq.\ (\ref{eq_wannier_ini}) better localizes
the wave function than just using $|\psi_{1\Vec{k}}\rangle, |\psi_{2\Vec{k}}\rangle$.
This is similar to monolayer graphene having the same honeycomb lattice structure,
where the superposition of the positive and negative energy states is required to have 
$A$-site or $B$-site localized orbital.


\begin{figure}
\begin{center}
\leavevmode\includegraphics[width=0.8\hsize]{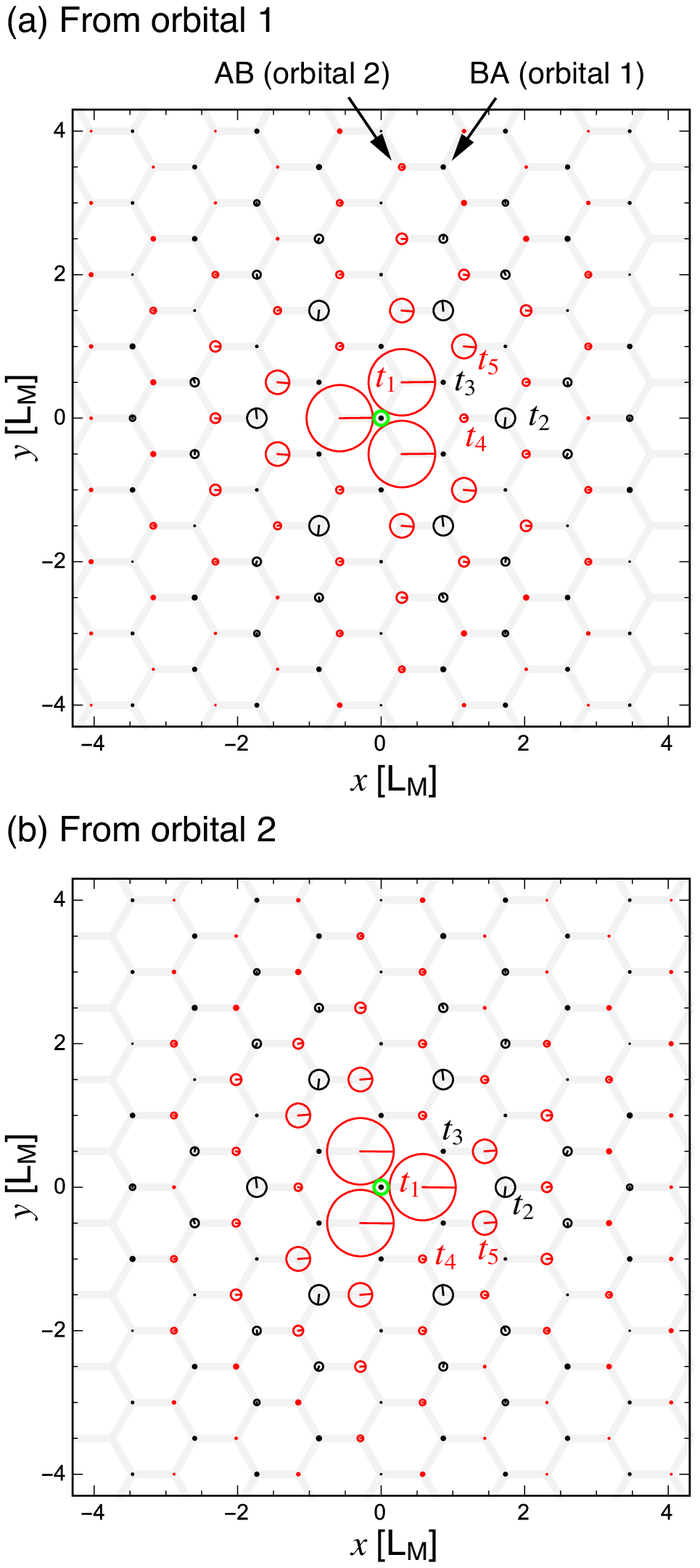}
\end{center}
\caption{
Hopping integrals in the effective tight-binding model 
for the low-energy flat band of TBG at $\theta=1.05^\circ$.
The panel (a) and (b) present the hopping parameters from the Wannier orbital 1 and 2,
respectively, where the radius of the circle at each lattice point
indicates the absolute value of the hopping integral from the origin to that point,
and the direction of the bar represents the phase in the complex plane.
}
\label{fig_hopping}
\end{figure}

\section{Effective tight-binding model}
\label{sec_tight-binding}

From the Wannier orbitals and the energy bands,
we can derive the effective tight-binding model to \textit{exactly} reproduce the dispersion of the nearly-flat bands.
In a straightforward calculation, the hopping integral between the Wannier orbitals is written as
\begin{align}
&
\langle \Vec{R}', n'| H| \Vec{R}, n \rangle \nonumber\\
&= 
\frac{1}{N} \sum_{\Vec{k}} e^{i\Vec{k} \cdot (\Vec{R}'-\Vec{R})}
\left[
\hat{U}^{(\Vec{k})\dagger}
\begin{pmatrix}
E_1(\Vec{k}) & 0 \\
0 & E_2(\Vec{k})  
\end{pmatrix}
\hat{U}^{(\Vec{k})}
\right]_{n'n},
\end{align}
where $\hat{U}^{(\Vec{k})}$ represents the matrix $U^{(\Vec{k})}_{mn}$, 
and we used 
$\langle \psi_{n'\Vec{k}'}| H| \psi_{n\Vec{k}} \rangle = \delta_{nn'}\delta_{\Vec{k}\Vec{k}'} E_n(\Vec{k})$.
In Fig.\ \ref{fig_hopping}(a) and (b), we plot the hopping integrals from orbital 1 and 2, respectively,
for the valley $\xi = +$.
Here the honeycomb lattice represents the network of BA spots (orbital 1) and AB spots (orbital 2).
The radius of the circle at each lattice point
indicates the absolute value of the hopping integral from the origin (green circle at the center) to that point,
and the direction of the bar represents the phase in the complex plane.
The effective tight-binding model for the valley $\xi = -$ is just given by taking the complex conjugate.
The list of the hopping integrals for the valley $\xi = +$ is included in Supplementary Material. \cite{suppl2}

To understand this effective tight-binding model, we need to analyze the symmetry properties of Wannier orbitals under point group $D_3$, as in Ref.\ \onlinecite{yuan2018model}. Recall that orbitals 1 and 2 have nonzero angular momentum $ L_{z}=-\xi $ at the valley $ \xi $. Furthermore, under two-fold rotation $C_{2y}$ which interchanges two graphene layers, we find orbital 1 from valley $\xi$ is mapped to orbital 2 from valley $-\xi$ and vice versa as shown in Fig.\ \ref{fig_wannier}. Hence we can regard orbitals 1 and 2 from valley $ \xi $ as the $ p $-wave-like orbitals $ p_{\xi}\equiv p_{x}+i\xi p_{y} $ residing on BA and AB spots respectively. The angular momentum of $ p_{\xi} $ orbital is $ L_{z}=-\xi $ whether its center is at BA or AB spot, which is consistent with $ L_{z} $ of orbital 1 and 2. Under $ C_{2y} $ the two graphene layers and hence BA and AB spots are interchanged, and $ (p_{x},p_{y})\to(-p_{x},p_{y}) $ or $ p_{\xi}\to -p_{-\xi} $. In other words, $ C_{2y} $ interchanges $ p_{\xi} $ orbital at BA spot and $ p_{-\xi} $ orbital at AB spot, which reproduces the symmetry transformation of orbital 1 and 2 under $C_{2y}$. 

Once we identify the symmetries of orbital 1 and 2, the tight-binding model then describes hopping among $ (p_x ,p_y) $ orbitals on the honeycomb lattice formed by BA and AB spots, which reads
\begin{eqnarray}\label{eq_tb0}
H=\sum_{\xi =\pm}\sum_{ij}{t}(\bm r_{ij})e^{i\xi\phi(\bm r_{ij})}c^\dagger_{i\xi}c_{j\xi},
\end{eqnarray}
where $ c_{i\xi} $ annihilates a $ p_{\xi} $-orbital electron at site $ i $, $ \bm r_{ij} $ is the vector from site $i$ to $j$, and $ t(\bm r),\phi(\bm r) $ are as shown in Fig. \ref{fig_hopping} (a) and (b). 

The symmetry group of the tight-binding model of Eq.\ (\ref{eq_tb0}) is $ G=D_{3}\times $U(1)$ \times $SU(2)$ \times T $, where $ D_{3} $ is the point group of TBG, which acts jointly on lattice sites and $ (p_x ,p_y) $ orbitals, U(1) acts in orbital space, SU(2) acts in spin space and $ T $ is the time-reversal symmetry. As discussed in Ref.\ \onlinecite{yuan2018model}, the microscopic origin of this orbital U(1) symmetry is that at small twist angles the intervalley coupling is strongly suppressed, leading to this approximate valley conservation that exists independent of crystal symmetries. 

\begin{figure}
\begin{center}
\leavevmode\includegraphics[width=1.\hsize]{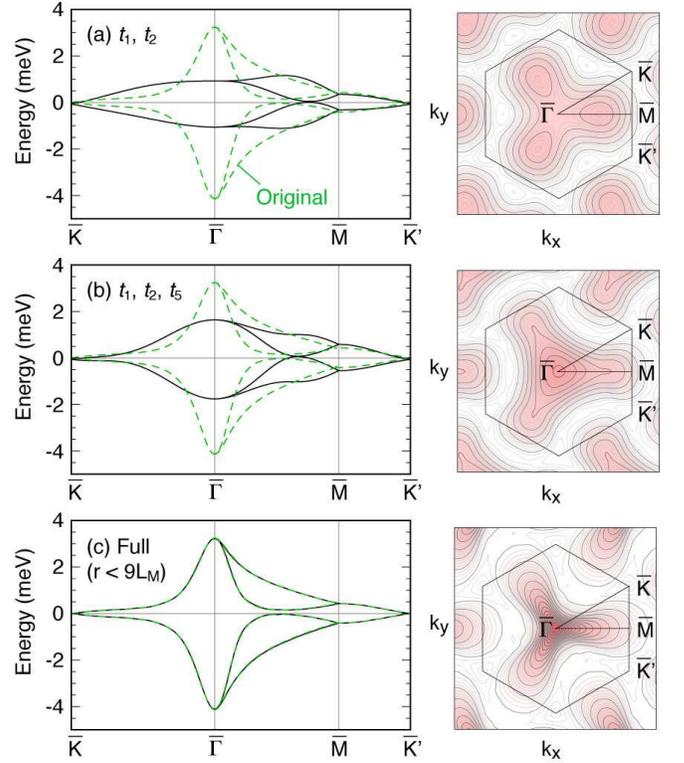}
\end{center}
\caption{Band structure in the effective tight-binding model for $\theta = 1.05^\circ$,
with (a) $t_1$ and $t_2$,  (b) $t_1, t_2$ and $t_5$
and (c) all the hopping parameters within the distance $r < 9L_{\rm M}$. 
Dashed line indicates the original energy band of the effective continuum model.
Right panels show the corresponding contour plots of  $E_1(\Vec{k})$ for the valley $\xi = +$. 
}
\label{fig_band_eff}
\end{figure}

The hopping integral $t(\bm r)$ roughly decays with increasing $r=|\bm r|$. To include dominant contributions, we consider the nearest five hopping integrals $ t_1 $ to $ t_5$ 
shown in Fig. \ref{fig_hopping} (a) and (b), which are within the range $ r\leqslant\sqrt{3}L_{\rm M} $. Notice that the subscript are not labeled according to $r$.
In the present model, we have 
$t_1\approx 0.331$ meV,  $t_2\approx (-0.010\pm 0.097i)$ meV,   $t_3\approx 0.016$ meV,
$t_4\approx  0.036$ meV, and $t_5 \approx 0.119$ meV. 
Figure \ref{fig_band_eff} presents the 
band structure in the effective tight-binding models  
with (a) $t_1$ and $t_2$,  (b) $t_1, t_2$ and $t_5$
and (c) all the hopping parameters within the distance $r < 9L_{\rm M}$. 
Dashed line indicates the original energy band of the effective continuum model.

With hopping terms $ t_{1} $ and $ t_{2} $ only, tight-binding model (\ref{eq_tb0}) becomes the minimum model introduced in Ref.\ \onlinecite{yuan2018model},
\begin{eqnarray}\label{eq_tb1}
H_{0}&=&-\mu\sum_{i}\bc^\dagger_{i}  \cdot\bc_{i}+\sum_{\langle ij\rangle}    t_1 \bc^\dagger_{i}  \cdot\bc_{j} +h.c.\\\nonumber
&+&\sum_{\langle ij\rangle'}\tilde{t}_2 \bc^\dagger_{i}  \cdot\bc_{j}+t_{2}'({\bc}^\dagger_{i}\times{\bc}_{j})_{z}+ h.c.
\end{eqnarray}
where $\bc_{i}= (c_{i,x}, c_{i,y})^{\text{T}}$ with $ c_{i,x(y)} $ annihilating an electron with $p_{x(y)}$-orbital at site $ i $, $ c_{j\xi}=(c_{jx}+i\xi c_{jy})/\sqrt{2} $. $ \mu $ is the on-site chemical potential, $ \tilde{t}_2={\rm Re}(t_2),t'_2={\rm Im}(t_2) $, and the sum over $ \langle ij\rangle' $ includes bonds with length $ \sqrt{3}L_{\rm M} $ along three directions $ \hat{\bm x},C_{3z}\hat{\bm x} $ and $ C_{3z}^2\hat{\bm x} $. The minimum tight-binding model (\ref{eq_tb1}) gives rise to a spectrum with Dirac nodes at $ \bar{K},\bar{K}' $ points. Notice that $ t_1 $ denotes hopping between two sublattices and we can always make $ t_1 $ real by properly choosing the relative phase between sublattices. The $ t'_{2} $ term describes the hexagonal warping effect in orbital space, which is responsible for band splittings along $ \bar{\Gamma}\bar{M} $ lines as shown in Fig. \ref{fig_band_eff}.

The symmetry group $G$ allows finite gaps at $ \bar{K},\bar{K}' $ points.\citep{yuan2018model}
However, due to the approximate sublattice symmetry at small twist angles,\citep{PhysRevB.86.155449} we can introduce an additional $ \mathbb{Z}_2 $ symmetry $ g:c_{{\bf R}\xi}\to c_{-{\bf R},-\xi} $, which combines twofold rotation in real space and chirality flip in orbital space. In the presence of $g$ and original symmetry group $G$, the gapless Dirac nodes at $ \bar{K},\bar{K}' $ points are guaranteed. The minimum model (\ref{eq_tb1}) satisfies both $G$ and $g$.
 
With this additional $\mathbb{Z}_2$ symmetry $ g $, we then consider additional hopping terms $ t_{3},t_{4} $ and $t_{5}$. As finite Im$(t_3)$ obeys $G$ while violates $ g $, we find Im$(t_3)=$0 from our numerical calculation of hopping integrals.
The nonzero $\tilde{t}_3\equiv{\rm Re}(t_3), t_{4},t_{5} $ terms preserve both $G$ and $g$, and quantitatively modify the band structure of the minimum model (\ref{eq_tb1}). In fact including $t_1$ to $t_5$ we have $ m_{e,h}^{-1}=3(\tilde{t}_{3}-3\tilde{t}_{2})\mp\left|\frac{1}{2}t_{1}+2t_{4}+7t_{5}\right| $ and $ v=\frac{\sqrt{3}}{2}\left|t_{1}-2t_{4}-t_{5}\right| $ where $ m_{e,h} $ denote effective masses at $ \bar{\Gamma} $ point on electron and hole sides respectively, and $ v $ is the Fermi velocity at $ \bar{K},\bar{K}' $ points.

We can also incorporate the effect of intervalley coupling in the effective tight-binding model by introducing U(1)-breaking hopping terms such as those in Ref.\ \onlinecite{yuan2018model}, which may explain Landau level degeneracy lifting in experiments.
A detailed analysis of Dirac nodes and mass generation will be presented in a forthcoming work.

\begin{figure}
\begin{center}
\leavevmode\includegraphics[width=0.7\hsize]{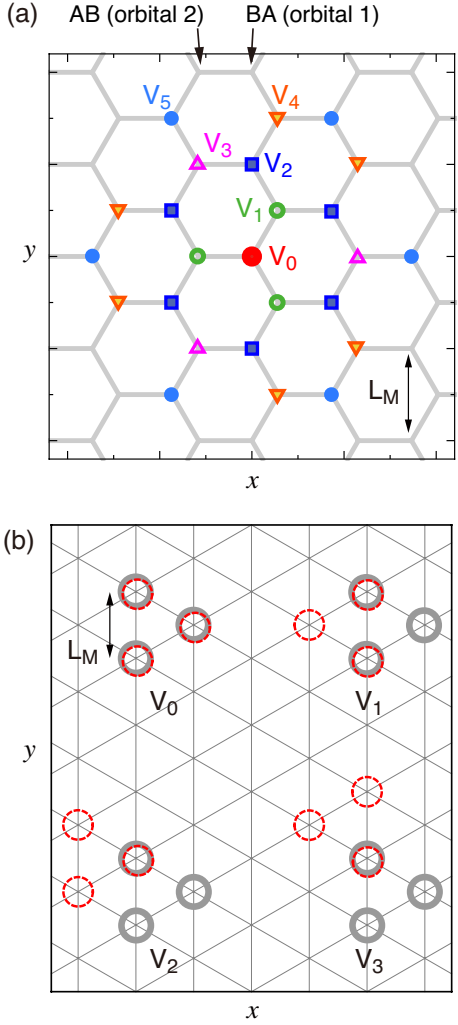}
\end{center}
\caption{(a) Labeling of the direct Coulomb interaction at different distances.
 $V_0, V_1, V_2 \cdots$ represent the potential amplitudes between the origin
and the indicated lattice points.
 (b) Overlapping of two Wannier orbitals in the configuration $V_0, V_1, V_2, V_3$.
The three circles of the same line type represent the three peaks of a single Wannier state
[Fig.\ \ref{fig_wannier}].
}
\label{fig_interaction}
\end{figure}

\begin{figure}
\begin{center}
\leavevmode\includegraphics[width=0.9\hsize]{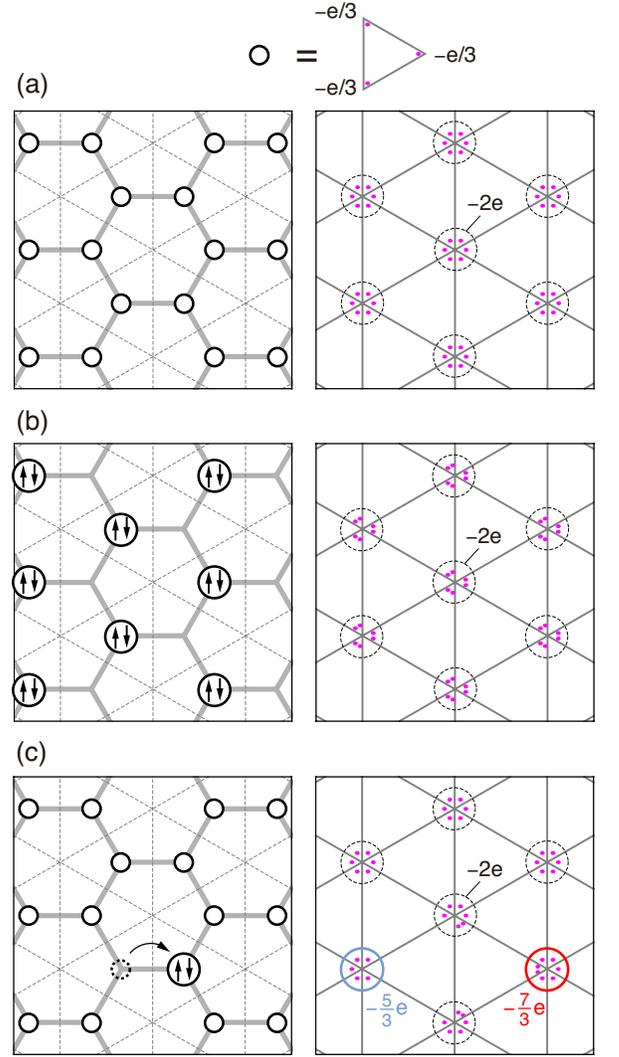}
\end{center}
\caption{
Several conceivable many-body states illustrated in the honeycomb lattice picture (left)
and the fractional charge picture (right).
(a) The homogeneous state where an electron resides at every sublattice.
(b) The charge-ordered state where two electrons enter every two sublattice.
Opposite arrows in a single site represent two electrons with different spins or valley pseudospins.
(c) An excitated state from the state (a) where an electron is transferred from a single site to another.
}
\label{fig_manybody}
\end{figure}

\section{Electron-electron interaction}
\label{sec_interaction}

We can calculate the electron-electron interaction parameters between the Wannier orbitals
directly from the wave functions obtained above.
The direct Coulomb interaction $V$ and the exchange interaction $J$
between $|\Vec{R}, m \rangle$ and $|\Vec{R}', m' \rangle$
are defined by
\begin{align}
&V_{\Vec{R}' m', \Vec{R} m} = \sum_{XX'} \iint d\Vec{r} d\Vec{r}'
|\psi^{X'}_{\Vec{R}' m'} (\Vec{r'})|^2 
\frac{e^2}{\epsilon |\Vec{r}-\Vec{r}'|}
|\psi^{X}_{\Vec{R} m} (\Vec{r})|^2,
\label{eq_direct}
\\
&J_{\Vec{R}' m', \Vec{R} m} = \sum_{XX'} \iint d\Vec{r} d\Vec{r}' \times
\nonumber\\
&\qquad\qquad
\psi^{X' *}_{\Vec{R}' m'} (\Vec{r}') \psi^{X *}_{\Vec{R} m} (\Vec{r}) 
\frac{e^2}{\epsilon |\Vec{r}-\Vec{r}'|}
\psi^{X}_{\Vec{R}' m'} (\Vec{r}) \psi^{X'}_{\Vec{R} m} (\Vec{r}'),
\label{eq_exchange}
\end{align}
where $\epsilon$ is the dielectric constant induced by the electrons in other bands
and by the external environment (e.g., the substrate).
The direct term is the classical Coulomb interaction and it works for any combinations of spin
and valley. On the other hand, the exchange interaction 
works only for the same spin and the same valley.
Rigorously speaking, the exchange term between different valleys (and the same spin)
is not exactly zero, but there the integral of
$e^{i(\Vec{K}_+ - \Vec{K}_-)\cdot (\Vec{r}-\Vec{r}')} /|\Vec{r}-\Vec{r}'|$ in Eq.\ (\ref{eq_exchange})
becomes much smaller than that for the same valley, so we neglect it.

\begin{table}
\caption{Direct interaction $V_n$ and the exchange interaction $J_n$
for the Wannier orbitals in units of $e^2/(\epsilon L_{\rm M})$.
The definition of $V_0, V_1\cdots $ is presented in Fig.\ \ref{fig_interaction}(a).
$V^{\rm (approx)}_n$ is the direct interaction terms estimated by the point-charge approximation (see the text).
}
\begin{center}
\begin{tabular}{c | cccccc}
$n$ & 0 & 1 & 2 & 3 & 4 & 5 \\
\hline
$V_n$ & 1.857 & 1.533 & 1.145 & 1.068 & 0.697 & 0.614 \\
$V^{\rm (approx)}_n$ & 1.857 & 1.524 & 1.136 & 1.081 & 0.679 & 0.610 \\
\hline
$J_n$ & N/A & 0.376 & 0.0645 & 0.010 & 0.014  & 0.001 
\end{tabular}
\end{center}
\label{tbl_int}
\end{table}

We label the direct interaction terms at different distances
as $V_0, V_1, V_2 \cdots$ as in Fig.\ \ref{fig_interaction}(a),
where $V_0$ is the on-site interaction, $V_1$ is the nearest neighbor interaction and so forth.
Similarly the exchange terms can be labeled as $J_1, J_2 \cdots$, where
$J_0$ does not exist due to the Pauli principle.
The calculated interaction parameters are listed in Table \ref{tbl_int}.
Here we notice that the on-site interaction $V_0$ is not much greater than others,
but it is in a similar magnitude to the nearest-neighbor interaction $V_1$.
The further interactions $V_2$ and $V_3$ are more than half of $V_0$.
This is quite different from usual Hubbard-type models where $V_0$ dominates the interaction effect. 
The peculiar distance dependence of Coulomb interaction in this model is closely related to 
the three-peak structure of the Wannier orbital.
For a single electron, each of three peaks accommodates the electric charge of $-e/3$,
and thus the interacting potential between two electrons can be written as a 
summation over the nine combinations of those fractional charges.
The direct Coulomb potential between two fractional charges located at the same peak
(i.e., the "on-site interaction" for the fractional charges) is $u_0 \approx (e/3)^2/\epsilon/(0.28 L_{\rm M})$,
while the potential between different peaks is well approximated by that for the point charges,
i.e., $(e/3)^2/\epsilon/r$, where $r$ is the distance between the peak centers.
The direct interaction terms estimated by this approximation are presented as $V^{\rm (approx)}_n$
in Table \ref{tbl_int}, where the error is found to be 1\% or less.

Obviously the dominant contribution to $V_n$ comes from the on-site part $u_0$.
As shown in Fig.\ \ref{fig_interaction}(b), the electrons located at the same orbital ($V_0$)
share all the three peaks, the nearest neighbor configuration ($V_1$) the two peaks,
and the next nearest ones  ($V_2, V_3$)  a single peak.
Therefore, the on-site interaction of the fractional charges
included in $V_0, V_1, V_2, V_3$ is $3u_0, 2u_0, u_0, u_0$, respectively,
and this explains the dominant part of the relative amplitudes of $V_n$'s.
These relatively long-range electron-electron interactions can potentially modify the hopping parameters and hence renormalize the low-energy band structure.

At the filling of two electrons per super cell, in particular, the triangular charge distribution results in 
an unexpected coincidence in the direct Coulomb energy between
two different many-body states shown in Fig.\ \ref{fig_manybody},
with (a) a homogeneous state where an electron resides at every sublattice of the honeycomb lattice,
and (b) a charge-ordered state where two electrons enter every two sublattice.
It may seem that the direct Coulomb energy in (b) is greater than in (a)
because of the double occupancy.
However, since an electron at the honeycomb site is actually composed of three 1/3 charges
at the triangle corners,
the states (a) and (b) have nearly identical charge distribution as shown in the right panels, where 
the charge of $-2e$ is registered to every AA spots.
Considering that the direct Coulomb interaction is very well approximated by the simple 
point-charge model as argued above,
the total direct energies of (a) and (b) must be nearly equal.
The competing nature of the two completely different states
may suggest a nontrivial many-body ground state.
For further consideration, we need to include
the exchange interaction and also the kinetic energy.
Lastly, Fig. \ref{fig_manybody}(c) illustrates an excitation from the state (a),
where an electron is transferred from a single honeycomb site to another.
This actually corresponds to a pair creation of the fractional charges $(\pm 1/3)e$ as shown
in the right panel. This is another intriguing property at this filling factor.

\section{Conclusion}
\label{sec_conclusion}

An extended Hubbard model is obtained for the nearly flat band in the low-angle TBG
by starting from the Bloch states in a realistic continuum model.
The obtained Wannier localized state is centered at AB or BA spot to form a honeycomb lattice.
The wave function of the Wannier orbital takes a triangular form which peaks at three AA spots surrounding the center,
and it leads to a competition between the on-site interaction and the neighboring interaction.
At the filling of two electrons per super cell, in particular, we have an unusual degeneracy of the 
a charge-ordered state and a homogeneous state, which implies an nontrivial nature of the ground state.
The detailed studies for the many-body ground states in this model will be left for future works.

\section{Acknowledgment}

MK thanks the fruitful discussions with Pilkyung Moon and Nguyen N. T. Nam. 
MK acknowledges the financial support of JSPS KAKENHI Grant Number JP17K05496. NFQY and LF are supported by the DOE Office of Basic Energy Sciences, Division of Materials Sciences and Engineering under award DE-SC0010526. LF is partly supported by the David and Lucile Packard Foundation.
TK is supported by JSPS KAKENHI Grant Number JP18K03442 and JST PRESTO Grant Number JPMJPR15N5.
KK is supported by JSPS KAKENHI Grant Number JP18H01860. 

{\it Note added:}
After completion of the present study, we have come to notice that
a recent preprint which also reports the maximally-localized Wannier states for TBG. \cite{kang2018symmetry}

\appendix

\section{Derivation of the effective continuum model under corrugation}
\label{sec_app}

We derive the effective interlayer interaction of a corrugated TBG in Eq.\ (\ref{eq_interlayer_matrix})
following the method in Ref.\ \onlinecite{moon2013opticalabsorption}.
We start from the single-orbital tight-binding model for $p_z$ orbital of carbon atoms.
We assume that the transfer integral between any two
orbitals is written
in terms of the Slater-Koster form as,
\begin{eqnarray}
&& -t(\Vec{R}) = 
V_{pp\pi}\left[1-\left(\frac{\Vec{R}\cdot\Vec{e}_z}{R}\right)^2\right]
+ V_{pp\sigma}\left(\frac{\Vec{R}\cdot\Vec{e}_z}{R}\right)^2,
\nonumber \\
&& V_{pp\pi} =  V_{pp\pi}^0 e^{- (R-a_0)/r_0},
\,\, V_{pp\sigma} =  V_{pp\sigma}^0  e^{- (R-d_0)/r_0}.
\label{eq_transfer_integral}
\end{eqnarray}
Here $\Vec{e}_z$ is the unit vector
perpendicular to the graphene plane,
$a_0 = a/\sqrt{3} \approx 0.142\,\mathrm{nm}$ is the distance of
neighboring $A$ and $B$ sites on graphene,
and $d_0 \approx 0.335\,\mathrm{nm}$
is the interlayer spacing of  graphite.
The parameter $V_{pp\pi}^0$ is the transfer integral between 
the nearest-neighbor atoms on graphene, and $V_{pp\sigma}^0$ is the transfer integral
between vertically located atoms on the neighboring layers of graphite. 
We take $V_{pp\pi}^0 \approx -2.7\,\mathrm{eV}$,
$V_{pp\sigma}^0 \approx 0.48\,\mathrm{eV}$, to
fit the dispersions of monolayer graphene.\cite{moon2013opticalabsorption} 
Here $r_0$ is the decay length of the transfer integral,
and is chosen as $0.184 a$ so that 
the next nearest intralayer coupling becomes $0.1 V_{pp\pi}^0$.

To construct the Hamiltonian matrix, we define the Bloch wave bases as
\begin{eqnarray}
&& |\Vec{k},A_l\rangle = 
\frac{1}{\sqrt{N}}\sum_{\Vec{R}_{A_l}} e^{i\Vec{k}\cdot\Vec{R}_{A_l}}
|\Vec{R}_{A_l}\rangle,
\nonumber \\
&& |\Vec{k},B_l\rangle = 
\frac{1}{\sqrt{N}}\sum_{\Vec{R}_{B_l}} e^{i\Vec{k}\cdot\Vec{R}_{B_l}}
|\Vec{R}_{B_l}\rangle,
\label{eq_bloch_base}
\end{eqnarray}
where 
the position $\Vec{R}_{A_l}(\Vec{R}_{B_l})$
runs over all $A(B)$ sites on the layer $l(= 1, 2)$,
$N$ is the number of monolayer's unit cell in the whole system,
and $\Vec{k}$ is two-dimensional Bloch wave vector 
defined in the first Brillouin zone of monolayer on the layer $l$.

For the interlayer coupling, we first consider a non-rotated bilayer graphene with $\theta = 0$
and a constant in-plane displacement $\GVec{\delta}$ from AA stacking.
The unit cell is spanned by  monolayer's 
lattice vectors, $\Vec{a}_1 = a(1,0)$ and $\Vec{a}_2 = a(1/2,\sqrt{3}/2)$,
which are now shared by both layers. 
Then the lattice points of the sublattice $X(=A_1,B_1, A_2, B_2)$ are given by
\begin{align}
&\Vec{R}_{A_1}=n_1\Vec{a}_{1}+n_2\Vec{a}_{2}+\GVec{\tau}_{A_1}, \nonumber\\
&\Vec{R}_{B_1}=n_1\Vec{a}_{1}+n_2\Vec{a}_{2}+\GVec{\tau}_{B_1}, \nonumber\\
&\Vec{R}_{A_2}=n_1\Vec{a}_{1}+n_2\Vec{a}_{2}+\GVec{\tau}_{A_2}+\GVec{\delta}+d(\GVec{\delta})\,\Vec{e}_z,
 \nonumber\\
&\Vec{R}_{B_2}=n_1\Vec{a}_{1}+n_2\Vec{a}_{2}+\GVec{\tau}_{B_2}+\GVec{\delta}+d(\GVec{\delta})\,\Vec{e}_z.
\end{align}
Here $\GVec{\tau}_{A_1}=\GVec{\tau}_{A_2} = 0$, 
$\GVec{\tau}_{B_1}=\GVec{\tau}_{B_2} = -\GVec{\tau}_1$ 
with $\GVec{\tau}_1 = (2\Vec{a}_2-\Vec{a}_1)/3 = (0, a/\sqrt{3})$,
and $d(\GVec{\delta})$ is the optimized interlayer distance which generally depends on $\GVec{\delta}$.
Note that $d(\GVec{\delta})$ is a periodic function of $\GVec{\delta}$ 
with periods of $\Vec{a}_{1}$ and $\Vec{a}_{2}$,
because the interlayer shift by a lattice vector just gives the equivalent structure.
It is known that the interlayer spacing takes
the maximum value $d_{\rm AA}$ at AA stacking ($\GVec{\delta}=0$),
and the minimum at $d_{\rm AB}$ at AB stacking ($\GVec{\delta}=\GVec{\tau}_1$) \cite{lee2008growth}.
Here we adopt $d_{\rm AA} = 0.360$ nm and $d_{\rm AB} = 0.335$ nm.\cite{lee2008growth,uchida2014atomic}
The distance at intermediate $\GVec{\delta}$ can be interpolated by
\begin{align}
d(\GVec{\delta}) = d_0 + 2d_1 \sum_{j=1}^3 \cos \Vec{a}^*_i \GVec{\delta},
\end{align}
with 
\begin{align}
&d_0 = \frac{1}{3} (d_{\rm AA} + 2d_{\rm AB}), \\
&d_1 = \frac{1}{9} (d_{\rm AA} - d_{\rm AB}).
\end{align}

The interlayer matrix element between from $X = A_1, B_1$ to $X' = A_2, B_2$
is obtained by taking all the transfer integrals between atoms of layer 1 and layer 2.
It is explicitly written as
\begin{align}
& U_{X'X}(\Vec{k},\GVec{\delta}) 
\equiv \langle \Vec{k},X' | H | \Vec{k},X\rangle \nonumber\\
&= \sum_{n_1,n_2}
- t[n_1 \Vec{a}_1 + n_2 \Vec{a}_2 + \GVec{\tau}_{X'X} + \GVec{\delta}+d(\GVec{\delta})\,\Vec{e}_z]
\nonumber\\
&\qquad \times
\exp\left[-i\Vec{k}\cdot(n_1 \Vec{a}_1 + n_2 \Vec{a}_2  + \GVec{\tau}_{X'X} + \GVec{\delta} )
\right],
\label{eq_interlayer_U}
\end{align}
where $\GVec{\tau}_{X'X} = \GVec{\tau}_{X'} - \GVec{\tau}_{X}$.
$U_{X'X}(\Vec{k},\GVec{\delta})$ is also a periodic function of $\GVec{\delta}$ 
with periods $\Vec{a}_{1}$ and $\Vec{a}_{2}$, and therefore
it can be written as a Fourier expansion, 
\begin{align}
& U_{X'X}(\Vec{k},\GVec{\delta}) 
\equiv \langle \Vec{k},X' | H | \Vec{k},X\rangle \nonumber\\
&= \sum_{m_1,m_2}
\tilde{U}_{X'X}(m_1 \Vec{a}^*_1 + m_2 \Vec{a}^*_2+ \Vec{k})
\nonumber\\
&\qquad \qquad \times
\exp\left[i(m_1 \Vec{a}^*_1 + m_2 \Vec{a}^*_2) \cdot( \GVec{\delta} + \GVec{\tau}_{X'X})\right].
\label{eq_interlayer_U2}
\end{align}
Here we defined
\begin{align}
&\tilde{U}_{X'X}(\Vec{q})
= -\frac{1}{S_0} \int t[\Vec{R} + d(\Vec{R}-\GVec{\tau}_{X'X})\,\Vec{e}_z] e^{-i\Vec{q}\cdot \Vec{R}} \, d^2\Vec{R},
\end{align}
where $S_0 = (\sqrt{3}/2)a^2$ is the unit area of monolayer graphene,
and the integral in $\Vec{R}$ is over the infinite two-dimensional space.
$\tilde{U}_{X'X}(\Vec{q})$ is circular symmetric and only depends on $|\Vec{q}|$.
Since $t(\Vec{R})$ exponentially decays in $R \, \sim \, r_0$, 
the Fourier transform $\tilde{U}_{X'X}(\Vec{q})$ decays in $q \, \sim \, 1/r_0$.

When we rotate one graphene layer to another by a small twist angle $\theta$,
the local lattice structure in the moir\'{e} pattern is approximately viewed as
a non-rotated bilayer graphene, where the displacement $\GVec{\delta}$ slowly depends on the position
$\Vec{r}$ in accordance with \cite{nam2017lattice}
\begin{align}
\GVec{\delta}(\Vec{r}) = [R(\theta/2)-R(-\theta/2)]\Vec{r}. 
\label{eq_delta}
\end{align}
The interlayer matrix element for valley $\xi$
is then approximately written by
$U_{X'X}[\Vec{K}_\xi,\GVec{\delta}(\Vec{r})]$.\cite{moon2013opticalabsorption} 
Using Eqs. (\ref{eq_interlayer_U2}) and (\ref{eq_delta}), we obtain
\begin{align}
& U_{X'X}[\Vec{K}_\xi,\GVec{\delta}(\Vec{r})]
\nonumber\\
&= \sum_{m_1,m_2}
\tilde{U}_{X'X}(m_1 \Vec{a}^*_1 + m_2 \Vec{a}^*_2+ \Vec{K_\xi})
\nonumber\\
&\qquad \times
\exp\left[i(m_1 \Vec{a}^*_1 + m_2 \Vec{a}^*_2) \cdot \GVec{\tau}_{X'X}\right].
\nonumber\\
&\qquad \times
\exp\left[i(m_1 \Vec{G}^{\rm M}_1 + m_2 \Vec{G}^{\rm M}_2) \cdot\Vec{r}\right],
\label{eq_interlayer_U3}
\end{align}
where we used the relationship 
$ \Vec{a}^*_i \cdot \GVec{\delta} (\Vec{r}) = \Vec{G}^{\rm M}_i \cdot \Vec{r}$.
Now we see that Eq.\ (\ref{eq_interlayer_U3}) is periodic in $\Vec{r}$
with the moir\'{e} reciprocal vectors $ \Vec{G}^{\rm M}_i$.
Since $\tilde{U}_{X'X}(\Vec{q})$ rapidly decays in $q$, 
we only need a few Fourier components in Eq.\ (\ref{eq_interlayer_U3}).
By taking the largest three terms given by $(m_1,m_2) = (0, 0), \xi(1, 0), \xi(1,1)$, 
we have the Hamiltonian in Eq.\ (\ref{eq_interlayer_matrix}), where
\begin{align}
& u = -\frac{1}{S_0} \int t[\Vec{R} + d(\Vec{R})\,\Vec{e}_z] e^{-i\Vec{K_\xi}\cdot \Vec{R}} \, d^2\Vec{R},
\nonumber \\
& u' = -\frac{1}{S_0} \int t[\Vec{R} + d(\Vec{R}-\GVec{\tau}_1)\,\Vec{e}_z] e^{-i\Vec{K_\xi}\cdot \Vec{R}} \, d^2\Vec{R}.
\end{align}
We obtain $u = 0.0797$eV and $u' = 0.0975$eV for the present parameters.
In a flat TBG, the interlayer distance $d(\GVec{\delta})$ is constant and therefore we have $u=u'$.\cite{moon2013opticalabsorption} 

\bibliography{wannier_TBG}

\end{document}